\documentclass[%
reprint,            
aps,
pre,
superscriptaddress,
floatfix
]{revtex4-2}

\usepackage{amsmath,amssymb,bm,mathtools}
\usepackage{subfig}

\usepackage{graphicx}     
\usepackage{caption}      

\usepackage{dcolumn}      
\usepackage{booktabs}     
\usepackage{bbm}

\usepackage[colorlinks=true,allcolors=black]{hyperref}

\begin{document}


\title{Random Walks Across Dimensions: Exploring Simplicial Complexes}

\author{Diego Febbe}
\email{diego.febbe@unifi.it}
\affiliation{Department of Physics and Astronomy, University of Florence \& INFN, Via Sansone
1, 50019 Sesto Fiorentino, Firenze, Italy}
\affiliation{Department of Mathematics \& Namur Institute for Complex Systems - naXys, University of Namur, Rue Joseph Grafé 2, 5000 Namur, Belgium}

\author{Duccio Fanelli}
\affiliation{Department of Physics and Astronomy, University of Florence \& INFN, Via Sansone
1, 50019 Sesto Fiorentino, Firenze, Italy}

\author{Timoteo Carletti}
\affiliation{Department of Mathematics \& Namur Institute for Complex Systems - naXys, University of Namur, Rue Joseph Grafé 2, 5000 Namur, Belgium}

\date{\today}

\begin{abstract}
We introduce a novel operator to describe a random walk process on a simplicial complex. Walkers are allowed to wonder across simplices of various dimensions, bridging nodes to edges, and edges to triangles, via a nested organization that hierarchically extends to higher structures of arbitrary large, but finite, dimension. The asymptotic distribution of the walkers provides a natural ranking to gauge the relative importance of higher order simplices. Optimal search strategies in presence of stochastic teleportation are addressed and the peculiar interplay of noise with higher order structures unraveled.  
\end{abstract}

\maketitle

\section{Introduction}
In the last few decades, Complex Networks (CNs) have proven essential in understanding a wide range of natural and technological problems, via captivating representations of the binary interactions among constituting units \cite{newman2010networks, barabasi2016network}. Examples range from biological applications (such as the patterns of interactions between proteins and/or genes) to neuroscience, via engineering related problems (as power grid stability and synchronization), socio-economic phenomena and the spreading of epidemics \cite{Barabasi2004, lucas2023inferring, barabasi2023neuroscience, nishikawa2015comparative, febbe2024chaos, 
buongiovanni2022will,
halaj2024financial,
pastor2001epidemic}. In all examined settings, non trivial interconnections between the observed dynamics and the underlying support emerge as solid evidence  \cite{boccaletti2006complex, delvenne2015diffusion, schaub2019structured}.

A further step in understanding the collective dynamics of systems composed of mutually entangled agents can be achieved by explicitly accounting for Higher-Order Interactions (HOIs). These latter are naturally incorporated in simplicial complexes and/or hypergraphs which thus define a novel modeling frontier to challenge the onset of emergent phenomena in complex environment populated by many agents in simultaneous interactions \cite{battiston2020networks, battiston2021physics,bick2023higher,Millan2025, peri2026spectral}. Notable applications of such generalized paradigm range from neuroscience to sociology, and potentially embrace a large gallery of distinct disciplinary realms   \cite{giusti2016two, petri2018simplicial, chowdhary2021simplicial, iacopini2019simplicial}. HOI models offer indeed an ideal platform to elaborate on the non trivial interplay between dynamics and topology, as already highlighted in \cite{Millan2025, muolo2024turing, gambuzza2021stability, gallo2022synchronization,Carletti2023}. Among the most fundamental dynamical processes that can be deployed on heterogeneous networks, and used to unravel the hidden features on the latter, random walks are worth mentioning. Random walks have been thoroughly studied on standard networks built on pairwise interaction \cite{lovasz1993random} and incessantly invoked as a fundamental modeling ingredient \cite{masuda2017random, bhattacharya2021random, Peri2026smart}. Starting from these premises, we here aim at filling an evident gap by extending the elemental random walks recipe to higher order simplices. At variance with past attempts \cite{carletti2020random, schaub2020random}, we here leverage on the generalized connectivity as naturally defined by the boundary operators, to drive topological hopping across simplices of contiguous dimensions. 
This scheme applies to settings relevant e.g. at the human scale, think for instance to information (or ideas) that can propagate from individuals to cliques and back again. 

Further applications arise in contexts where dynamics evolve on and are shaped by higher-order interactions. In neuroscience, the brain is increasingly modeled as a simplicial complex, where neurons (0-simplices) are connected via synapses (1-simplices) and influenced by higher-order interactions composed by neural cliques of co-firing neurons \cite{giusti2016two, reimann2017cliques, petri2014homological, bassett2017network}. \\
Moreover, in protein or gene regulation interactions often involve more than pairwise relationships, motivating descriptions in terms of simplicial complexes or hypergraphs \cite{malod2019functional, camara2017topological, wei2022hodge}. In this context, a dynamical process that transitions across dimensions, such as among higher-order functional clusters, provides a natural framework to model how signals propagate between different topological scales.\\
More generally, the proposed cross-dimensional random walk can be interpreted as a topological probing mechanism for empirical complex systems, including social, biological, and technological networks enriched with clique complexes. In such settings, it enables the community detection of higher-order organization beyond dyadic interactions. In particular, communities are not only characterized by dense edge connectivity, but also by regions enriched in triangles, tetrahedra and higher-dimensional simplices, creating functional topological modules \cite{benson2016higher, lambiotte2015random}.

Working within the proposed scheme we will show that the asymptotic distribution of the walkers returns a sensible ranking for the relative importance of the simplices. Noise - acting as a form of long range teleportation - is also introduced and shown to significantly potentiate the ability of the walkers to navigate within a complex environment decorated with higher order structures, more effectively than what it happens for the reference case where agents are solely bound to hop across vertices. In the following we will begin by defining the suggested mathematical model.

\section{The model}
\begin{figure}[h!]
    \centering
    \includegraphics[width=0.9\linewidth]{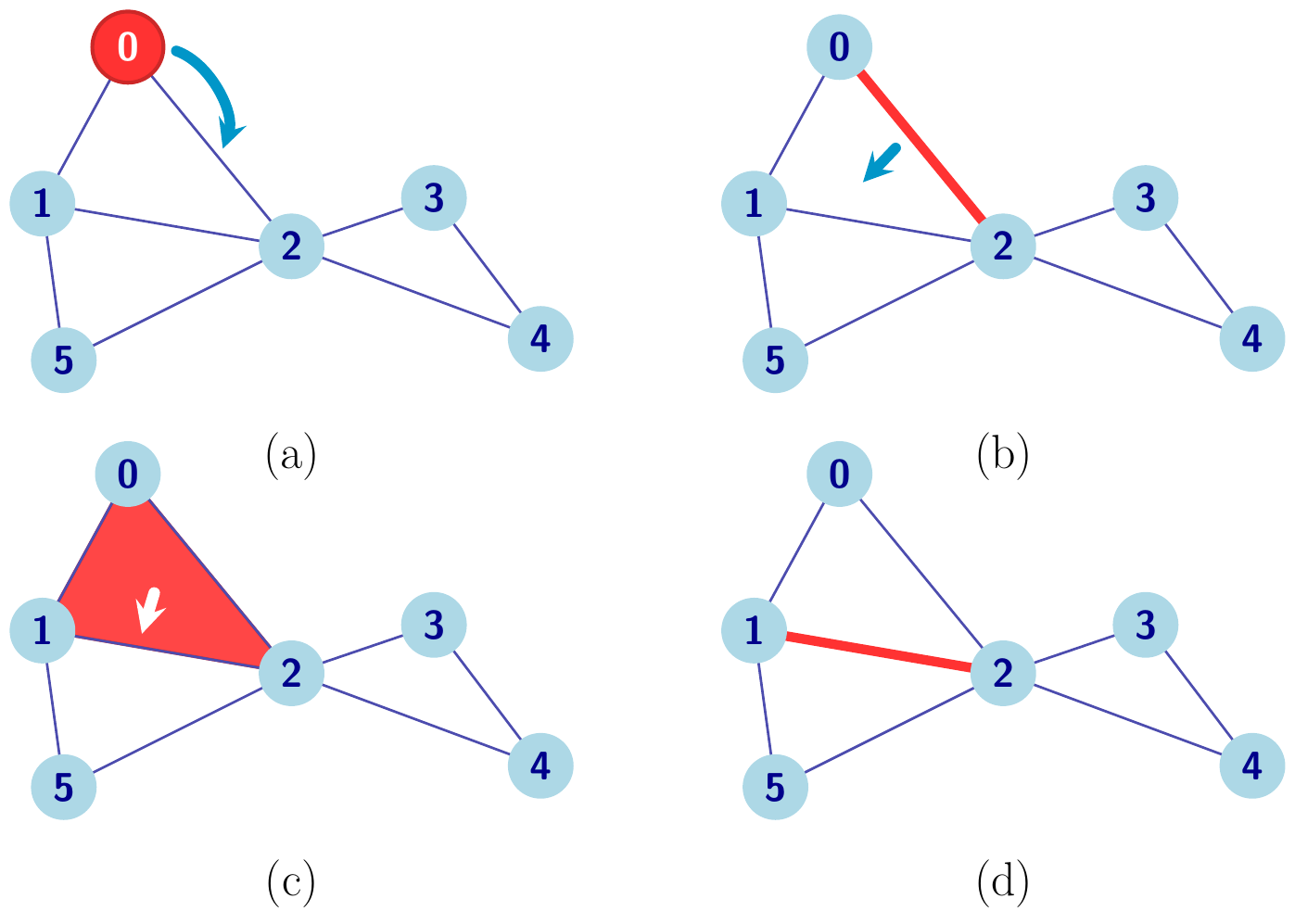}
    \caption{Example of a random walk process across the dimensions of a simplicial complex. Here, the walker is initialized on the node $0$ (panel (a)). Then, it reaches the connected link $[0,2]$ (panel (b)), heads towards the triangle $[0,1,2]$ (panel (c)) and finally lands on link $[1,2]$ (panel (d)).}
    \label{fig:example_of_RW_on_SC}
\end{figure}

In Fig. \ref{fig:example_of_RW_on_SC} we provide a pictorial representation of the process that we set to model and that is ultimately intended to reproduce the dynamics of a walker wandering across the simplicial complex dimensions. In the displayed cartoon, the walker is initialized on a node, panel (a). Then, in the next step of evolution, it relocates to eventually land on one of the connected simplices (a link), see panel (b). From there, the process is iterated forward and the walker first reaches an incident triangle - panel (c) - before transitioning towards one of the edges, that delimits the boundary of the departure face.

To mathematically tackle the above process, we begin by considering a $D$-simplicial complex, $\mathcal{X}$, containing $N_k$ $k$-simplices, $0\leq k \leq D$, $\sigma_i^{(k)}$, $i=1,\dots,N_k$, the latter being composed by $(k+1)$ nodes, i.e., $0$-simplices. Once an orientation has been set for the simplices, the simplicial complex is completely defined by the incidence matrices $\mathbf{B}_k\in \mathbb{R}^{N_{k-1}\times N_k}$, $k=1,\dots,D$, that set the underlying topological structure, namely how simplices are {\em assembled} together, by taking into account their relative orientation~\cite{grady2010discrete,bianconi2021higher}. We shall use in particular the {\em unsigned} incidence matrices $\mathbf{A}_k\in \mathbb{R}^{N_{k-1}\times N_k}$, whose entries equal $1$ if the $(k-1)$-simplex is incident, i.e., it is a face, of the $k$-simplex, and $0$ otherwise. In formulae:
\begin{equation}
\label{eq:Mk}
A_k(\sigma_i^{(k-1)},\sigma_j^{(k)})=1 \text{ iff } \sigma_i^{(k-1)}\subset \sigma_j^{(k)}\, .
\end{equation}
For $k=1,\dots, D$, we can introduce the ``upper'' degree of the $(k-1)$-simplex, $\sigma_i^{(k-1)}$, namely the number of $k$-simplices to which it belongs to, by
\begin{equation}
\label{eq:dsigmaup}
d_i^{(k-1)} = \sum_{j=1}^{N_k} A_k(\sigma_i^{(k-1)},\sigma_j^{(k)})\, .
\end{equation}
In the case $k=1$, i.e., the simplex $\sigma_i^{(0)}$ is a node, then $d_i^{(0)}$ returns the standard (network) node degree. Observe that we can set $d_i^{(D)}=0$ for all $i=1,\dots,N_D$, because in the $D$-simplicial complex, by definition all the maximal simplices $\sigma_i^{(D)}$ are not faces of any larger simplices.

We can also introduce the ``lower'' degree of the $k$-simplex, $\sigma_j^{(k)}$, $k=1,\dots, D$, namely the number of $(k-1)$-simplices contained into it; however because of the  closure condition of simplicial complexes, we can trivially state
$\delta_j^{(k)} = \sum_{i=1}^{N_{k-1}} A_k(\sigma_i^{(k-1)},\sigma_j^{(k)})\equiv k+1 $. Let us observe that since nodes do not accommodate for smaller simplices, $\delta_j^{(0)}=0$, for all $j=1,\dots,N_0$. As stated above, hops occurs between incident or adjacent simplices, namely only steps towards elements characterized by one dimension higher or lower are allowed. Further, we
assume time to evolve by discrete steps, i.e., each time step the walker performs a hop from one simplex to another one. We define $p_t(\sigma_i^{(k)})$ the probability to find the walker at time $t$ on the simplex $\sigma_i^{(k)}$ and assume an unbiased random walk. For all $0<k<D$ the master equation ruling the time evolution of the above defined probability reads:
 \begin{equation}
\label{eq:ptk}
\begin{aligned}
p_{t+1}(\sigma_i^{(k)}) = &\sum_{j=1}^{N_{k-1}} p_{t}(\sigma_j^{(k-1)})\frac{A_k(\sigma_j^{(k-1)},\sigma_i^{(k)})}{d_j^{(k-1)}+\delta_j^{(k-1)}}+\\
&\sum_{l=1}^{N_{k+1}} p_{t}(\sigma_l^{(k+1)})\frac{A_{k+1}(\sigma_i^{(k)},\sigma_l^{(k+1)})}{d_l^{(k+1)}+\delta_l^{(k+1)}}\, ,
\end{aligned}
\end{equation}
where the leftmost term on the right hand side stands for hops from $(k-1)$ to the $k$-simplices, while the rightmost one to hops from a $(k+1)$-to the $k$-simplices. The special cases $k=0$ and $k=D$ are given by
\begin{eqnarray}
\label{eq:pt0} 
p_{t+1}(\sigma_i^{(0)}) &=& \sum_{j=1}^{N_{1}} p_{t}(\sigma_j^{(1)})\frac{A_{1}(\sigma_i^{(0)},\sigma_j^{(1)})}{d_j^{(1)}+2}, \\
p_{t+1}(\sigma_i^{(D)}) &=& \sum_{j=1}^{N_{D-1}} p_{t}(\sigma_j^{(D-1)})\frac{A_D(\sigma_j^{(D-1)},\sigma_i^{(D)})}{d_j^{(D-1)}+D} \nonumber .
\end{eqnarray}

Introduce for $k=0,\dots, D$, the vectors $\vec{p}_t^{(k)}=(p_t(\sigma_1^{(k)}),\dots, p_t(\sigma_{N_k}^{(k)}))\in \mathbb{R}^{N_k}$, and the {\em transition} matrices $\hat{\mathbf{M}}_k$ and $\tilde{\mathbf{M}}_k$ whose elements are respectively $\hat{M}_k(j,i) =\frac{A_k(\sigma_j^{(k-1)},\sigma_i^{(k)})}{d_j^{(k-1)}+\delta_j^{(k-1)}}$ and $\tilde{M}_{k+1}(i,l)=\frac{A_{k+1}(\sigma_i^{(k)},\sigma_l^{(k+1)})}{d_l^{(k+1)}+\delta_l^{(k+1)}} ,$
then, the above process can be recast in the more compact form:

\begin{equation}
\label{eq:pt}
\vec{p}_{t+1}=\vec{p}_t\mathbf{M}\, .
\end{equation}

where $\vec{p}_t$ is the (row) vector $(\vec{p}_1^{(0)},\dots,\vec{p}_t^{(D)})\in\mathbb{R}^N$ and $\mathbf{M}\in \mathbb{R}^{N\times N}$, with $N=N_0+\dots+N_D$ reads
\begin{equation}
\label{eq:MatrixM}
\mathbf{M}=\left(
\begin{matrix}
 \mathbf{O}_{0} & \hat{\mathbf{M}}_1 & \dots & \dots & \dots & \dots& \dots\\
\tilde{\mathbf{M}}_1^\top & \mathbf{O}_{1} & \hat{\mathbf{M}}_2 & \dots & \dots & \dots& \dots\\
\vdots & \tilde{\mathbf{M}}_2^\top & \mathbf{O}_{2} & \hat{\mathbf{M}}_3 & \dots & \dots& \dots\\
\vdots &\vdots & \ddots &\ddots &\ddots & \dots& \dots\\
\vdots &\vdots & \vdots &\tilde{\mathbf{M}}_k^\top & \mathbf{O}_{k} & \hat{\mathbf{M}}_{k+1}& \dots\\
\vdots &\vdots & \vdots &\vdots & \ddots &\ddots& \ddots&\\
\end{matrix}\right)\, ,
\end{equation}

$\mathbf{O}_{k}$ being the null $N_k\times N_k$ matrix. It is straightforward to prove that the above discrete time dynamical system~\eqref{eq:pt} preserves the total probability, namely
$\sum_{k=0}^D\sum_{i=1}^{N_k} p_{t}(\sigma_i^{(k)})=1 \quad \forall t\geq 0$.
Hence, the largest eigenvalue of the stochastic matrix $\mathbf{M}$ is $\lambda_{\max}=1$ and it is associated to the normalized eigenvector 

\begin{equation}
\begin{aligned}
    \vec{u} = \frac{1}{\hat{N}}\bigl(
    &d_1^{(0)}, \dots, d_{N_0}^{(0)},\,
    d_1^{(1)}+2, \dots, d_{N_1}^{(1)}+2,\,
    \dots,\, \\
    &d_1^{(D-1)}+D, \dots, d_{N_{D-1}}^{(D-1)}+D,\,
D+1, \dots, D+1
\bigr)
\end{aligned}
\end{equation}
 with 
 
 \begin{equation}
 \begin{aligned}
     \hat{N}:=&2N_1+\dots+(D+1)N_D+\sum_{k=0}^{D-1}\sum_{i=1}^{N_k}d_i^{(k)}\\ =&\sum_{k=1}^D2(k+1)N_k.
 \end{aligned}
 \end{equation} 
 The vector $\vec{u}$ represents the asymptotic occupation probability as obtained by iterating forward the dynamics via map (\ref{eq:pt}). 

The cross-dimensional dynamical process expressed by the operator shown in Eq. \eqref{eq:MatrixM} can further enriched with jumps across various dimensions. We elaborate on that in the Appendix.

To challenge numerically the dynamical scheme introduced above, we need to preliminary define a suitable algorithm for simplicial complex generation. In the following, we 
briefly comment on the solution that we have adopted and which represents the ideal extension of the preferential attachment mechanism to higher order graphs. In particular, at each generation  step:

\begin{itemize}
    \item a new node enters the structure; 
    \item at this point, the dimension $k+1$, of the simplex to be formed, is chosen with probability $p_{k+1}$, with $\sum_{k=0}^{D-1}p_{k+1}=1$;
    \item then, the simplex $\sigma_j^{(k)}$ of dimension $k$ is selected with probability $\Pi_j^{(k)} = \Delta_j^{(k)}/\sum_{h=1}^{N_k} \Delta_{h}^{(k)}$. Here, $\Delta^{(k)}$ stands for a proxy of the degree which can be set either to $d^{(k)}+\delta^{(k)}$, or as we chose to do in this case, to $d^{(k)}$ so as to avoid dimensional biases. The selected simplex $\sigma_j^{(k)}$ is the module to which the incoming node attaches to create a new simplex of dimension $k+1$ with all the sub-simplices required by the closure property.
\end{itemize}

In the experiments reported throughout this work, we have chosen to solely adjust the triplet $(p_1,p_2,p_3)$, under the normalization condition $p_1+p_2+p_3=1$. Hence, all $p_k$ with $k>3$ are identically equal to zero and $D=3$. Further details on the generation algorithm are provided in \cite{febbe2026simplicial}. As a side comment, it is worth stressing that the above scheme allows to deal with simplexes of different dimensions and aligns with  established mechanisms for simplicial complex building that are, however, limited to consider simplexes of a given dimension at the time: by setting $p_D=1$, and $\Delta^{(k)}=d^{(k)}$ one recovers the framework presented in \cite{bianconi2016network} and analyzed in \cite{bianconi2017emergent, torres2020simplicial}, with the so called \textit{flavor} value set to unit. Other examples of methods for SC generation are reported in \cite{zuev2015exponential, dorogovtsev2025deterministic}, while, in parallel, a comprehensive collection of real-world datasets is provided in \cite{ahornDataset}.

\section{Higher Order Ranking}
\begin{figure}[h!]
    \centering
    \includegraphics[width=0.99\linewidth]{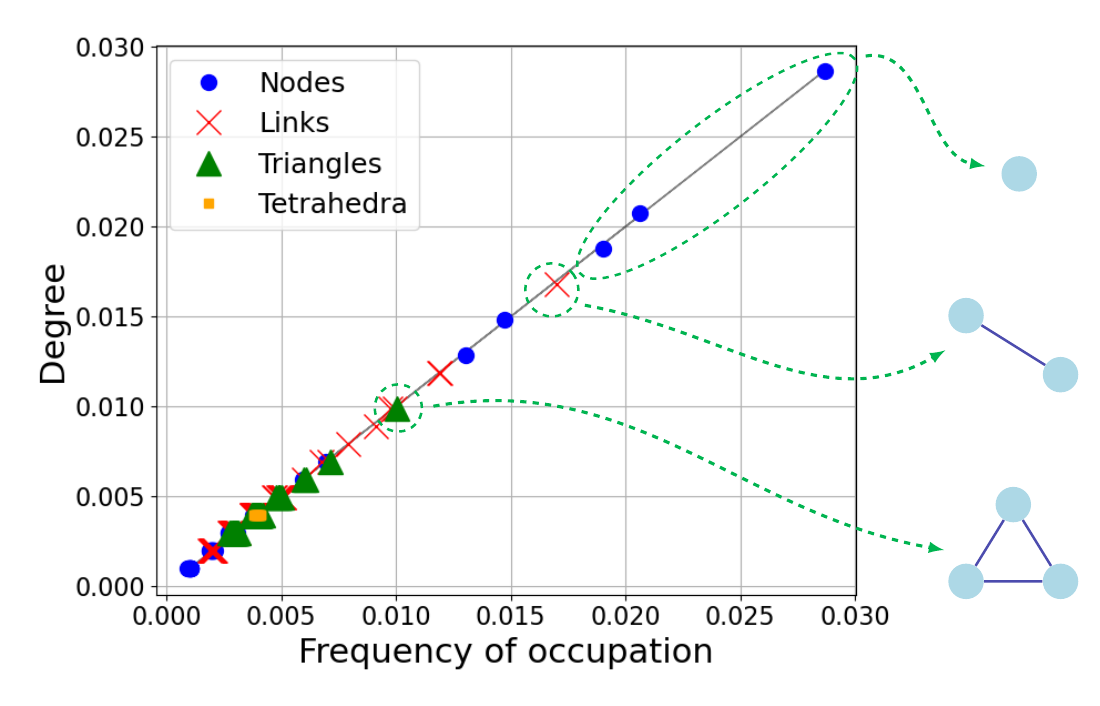}
    \caption{Correspondence between the asymptotic frequency of occupation on the simplices (symbols) and the corresponding generalized degree (solid line). The degree centrality, corresponding to the occupation probability of the simplices, naturally mixes the various dimensions. Here we set, $p_1 = p_2 = p_3 = 1/3$ and $N_0 = 50$.}
    \label{fig:mixed_ranking}
\end{figure}

Random walks evolving on a higher order graphs of the type generated above and following the prescriptions of the dynamics, as ruled by equation (\ref{eq:pt}),  distribute asymptotically across available dimensions - nodes, links, triangles, for the case here addressed - with a population fraction that reflects the associated generalized degree $\Delta_i^{(k)} = d^{(k)}_i+\delta^{(k)}_i$.
This latter quantity (or, equivalently, the measured density of walkers on each sub-structure, at equilibrium) provides an indirect measure of the relative importance of fundamental simplices, across their hosting simplicial complex. The ensuing ranking, as induced by the operator $\mathbf{M}$, naturally spans across all dimensions, mixing higher- and lower-order structures. Nodes can for instance pop out as most relevant structures, followed by links and triangles, with the interposed presence of additional nodes, as for displayed in Fig. \ref{fig:mixed_ranking}. Symbols stands for direct simulations of the microscopic dynamics (an ensemble made of non interacting agents wonder across the dimension of the generated simplex), while the solid line refers to the theoretical solution, as discussed above (the normalized eigenvector of the stochastic matrix $\mathbf{M}$, relative to its largest eigenvalue). Tracking generalized random walkers through time allows us to weight the relative importance of $k$-dimensional structures of different dimensions and types, extending beyond the limited perspective of $0$-dimensional units (the nodes), which define the restricted basin of exploration of their classical counterparts.

\section{Explore assisted by noise}
As a next step in the story, we modify the dynamics by accounting for the possibility of stochastic teleportation. More specifically the dynamics will combine local random walks to long-range jumps towards distant connected structures. Mathematically, this amounts to impose:
\begin{equation}
\label{eq:ptalpha_sparse}
\vec{p}_{t+1}=\vec{p}_t\left(\alpha \mathbf{M}+(1-\alpha) D_s\mathbf{S}\right), 
\end{equation}
where $\mathbf{S}$ is a random symmetric sparse matrix, with density 
$\delta = \sum_{i,j}S_{ij}/N^2$, and $D_s=\text{diag}(\vec{k_s})^{-1}$ with $k_s=\sum_j S_{ij}$, (see \cite{di2015optimal}). It is worth remarking that matrix $\mathbf{S}$ bridges simplices of any dimension, thus overcoming the topological constrains that are intrinsic to the proposed random walk recipe. Further, note that the case $\delta=1$ constitutes a generalization of the    celebrated PageRank algorithm \cite{page1999pagerank} to the analyzed setting where higher order structures are accounted for. The parameter $\alpha$ weights the relative importance of the two combined effects. When $\alpha=1$ the systems performs a pure random walk, and the agent can relocate only by engaging only incident simplices according to the underlying structure. Conversely, when $\alpha=0$, random jumps are 
the sole allowed modality to displace in space. Because of the stochastic distribution of the teleportation site, this latter mechanism is conceptualized as a source of noisy perturbation. It is well known that by properly tuning a stochastic contributions of the type introduced (via adjusting the control strength $\alpha$) results in optimal search strategies throughout binary networks \cite{di2015optimal}. The First Passage Time $T_{ij}$, the average number of steps that the walker has to take to reach simplex $j$, starting from simplex $i$, is the quantity to be monitored, as a function of $\alpha$, to shed light onto this intriguing phenomenon. Hereafter, we will restrict the starting and final state to $0$-dimensional simplices (i.e., the nodes), to enable for a direct comparison with what is obtained by operating on ordinary graphs. Following \cite{di2015optimal}, we can analytically compute the terms $T_{ij} = \sum_{h=1}^{N-1}\left(\mathbf{Z_j}^{-1}\right)_{ih}$
where $\mathbf{Z_j} = \mathbb{I}-\mathbf{P}_j$, with the node's index $j$ identifying the sub-matrix obtained from $\mathbf{P}$ by removing its $j$-th row and column. Here, $\mathbf{P} =\alpha \mathbf{M}+(1-\alpha) D_s\mathbf{S}$. By averaging $T_{ij}$ over both indices $i$ and $j$ such that $1\leq i,j\leq N_0$ (namely all possible departure and arrival nodes), we obtain a scalar measure of the node \textit{explorability}
\begin{equation}
    \langle T \rangle = \frac{1}{N_0(N_0-1)}\sum_{1\leq i\ne j \leq N_0}  T_{ij}.
    \label{eq:explorability_di_patti}
\end{equation}

\begin{figure} [h!!]
    \centering
    \subfloat[]{\includegraphics[width=0.46\linewidth]{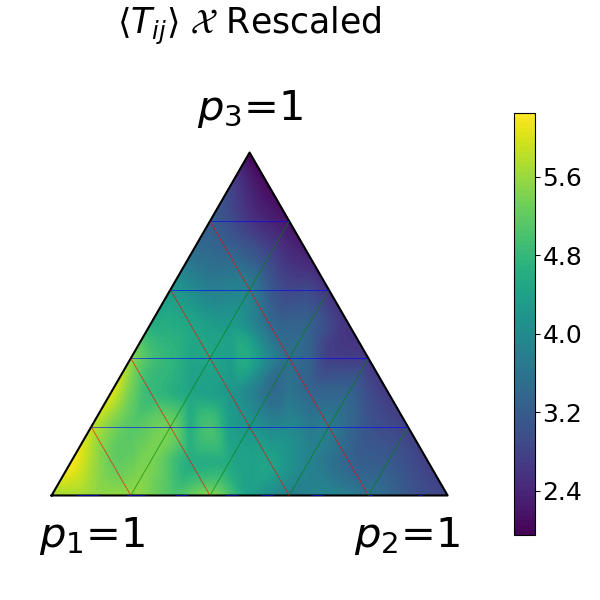}}
    \subfloat[]
    {\includegraphics[width=0.46\linewidth]{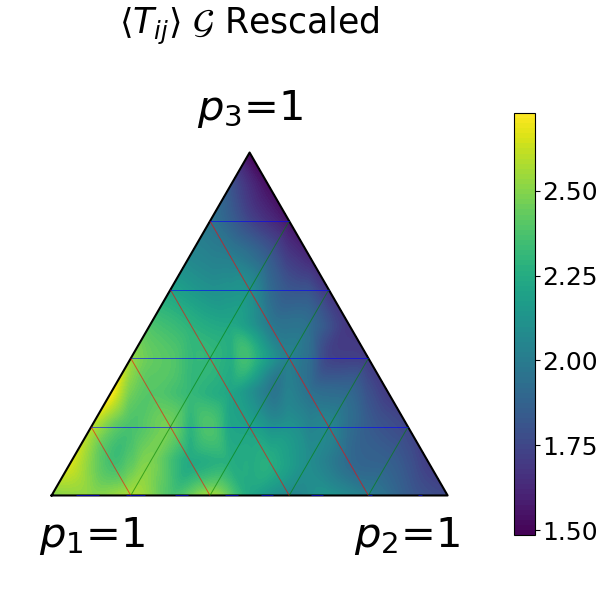}}
    \caption{Each point of these triangles corresponds to a choice of the parameter probabilities defining the topology of the generated simplex $\mathcal{X}$.  Panel (a): Mean node FPT for the generated simplicial complex divided by the total number of simplices. Panel (b): Mean FPT for $\mathcal{G}$, divided by the number of nodes (the sub-structures on which the walker is confined, in this case), fixed at $N_0=20$.}
    \label{fig:FPT_simplex_graph_normalized}
\end{figure}

Notice that given a simplicial complex $\mathcal{X}$, one can always define the underlying binary network $\mathcal{G}$ by just extracting the node-to-node interactions encoded by the links. We can therefore apply Eq.  (\ref{eq:explorability_di_patti}) to the
higher order support $\mathcal{X}$, on the one side, and to its corresponding graph $\mathcal{G}$, on the other, so as to compare the respective performance in terms of explorability, the ability to effectively explore the spatial support to which the system is eventually bound. For a fair comparison, the estimated average time $\langle T \rangle$ should be normalized by the extensive parameter that measures the total number of individual structures that the walker can visit when wandering across dimensions. These are the number of nodes, for $\mathcal{G}$, and the total number of simplices for 
$\mathcal{X}$. We begin by analyzing the setting $\alpha=1$, which corresponds to dealing with pure diffusion. The results are reported in Fig. \ref{fig:FPT_simplex_graph_normalized} where the explorability measure $\langle T \rangle$ (normalized by the total number of structures, for either $\mathcal{X}$ and $\mathcal{G}$) is plotted 
as a function of the generative parameters $p_1, p_2, p_3$, under the constraint $p_1 +p_2 +p_3 = 1$ imposed by the constructive algorithm. Apart from a slight difference (see annexed scales) which seems to indicate that the average search time on a simplex is further slowed down, besides obvious extensive scaling, the displayed patterns are pretty similar.

\begin{figure*}[t]
\centering

\refstepcounter{figure}\label{fig:gain_simplex_graph_summary}

\raisebox{-0.7cm}{%
\begin{minipage}[t]{0.55\textwidth}
  \centering

  \subfloat[]{%
    \includegraphics[width=0.9\linewidth]{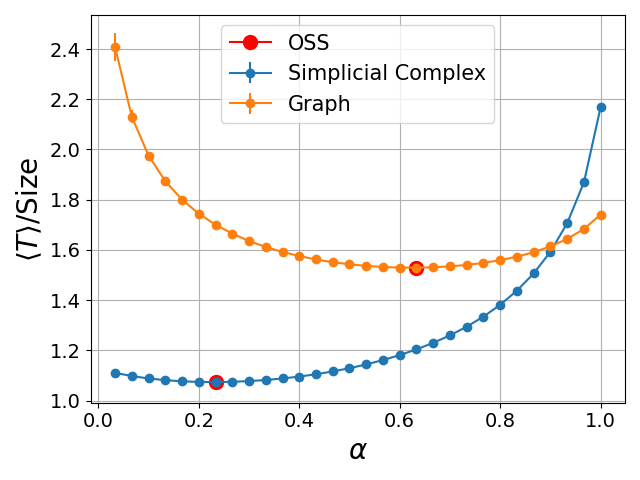}
  }
  {\captionsetup{type=figure,labelformat=empty,labelsep=none}
  \caption*{\text{\figurename~\thefigure.} %
  Panel (a): Size-rescaled explorability for a simplicial complex and its corresponding graph, as a function  of $\alpha$. The red dot identifies the minimum of the reported curves, namely the optimal search strategies (OSS). Here, we set $p_3=1$ and $\delta=0.05$.  Panel (b): Structural gain $g_s = \frac{g^\mathcal{X}}{g^\mathcal{G}}$, computed as the ratio between the gain of the simplicial  complex topology and that of the corresponding graph topology, for $\delta=0.5$. Panel (c) $g_s$ as a function of $\delta$. The shaded region traces the variability across different realizations of the generated simplices.}}
\end{minipage}%
}
\hfill
\begin{minipage}[t]{0.4\textwidth}
  \centering
  \subfloat[]{%
    \includegraphics[width=\linewidth]{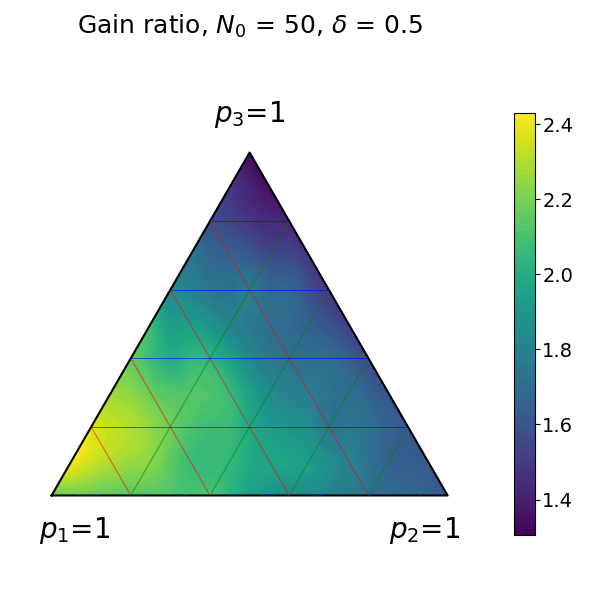}
  }
  \vspace{-0.3cm}
  \subfloat[]{%
    \includegraphics[width=\linewidth]{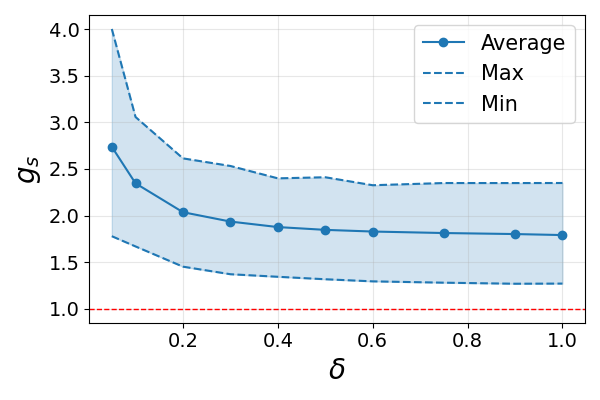}
  }
\end{minipage}
\end{figure*}

A radical different scenario is found for $\alpha<1$, i.e. when noise is thus allowed for. For any given topological support, the computed value of $\langle T \rangle$, also rescaled, is now influenced by two concurrent dynamical processes: local search and long-ranged teleportation. In Fig. \ref{fig:gain_simplex_graph_summary} we report on a typical outcome of the analysis for this generalized setting. In panel (a) the properly normalized time $\langle T \rangle$ is plotted as a function of $\alpha$, for a specific choice of the construction parameters $p_1$, $p_2$, $p_3$. A minimum is clearly displayed for both depicted curves, pointing to the existence of an optimal choice of the control parameter $\alpha$ that gauges the relative balance of the two combined dynamical effects. For large values of $\alpha$, indicating a predominance of local random walk, the size-normalized explorability $\langle T \rangle$, is larger for simplicial complexes as compared to their respective underlying graph, in agreement with the results reported in Fig. \ref{fig:FPT_simplex_graph_normalized}. Nevertheless, by progressively strengthening the teleportation term, we observe a more pronounced drop of the (rescaled) search time computed for the simplicial complex. Consequently, the corresponding gain in term of (noise assisted) exploration capability $g^a = \frac{\langle T \rangle(\alpha=1)}{\min_\alpha(\langle T \rangle)}$, for $a=\mathcal{X}$ or $\mathcal{G}$, is larger for the simplicial complex $\mathcal{X}$, when compared to what is found on the corresponding graph $\mathcal{G}$, see Fig. \ref{fig:gain_simplex_graph_summary} panel (b) for a specific choice of $\delta$. Simplicial complexes prove thus more responsive to the beneficial effects induced by the imposed stochasticity for an effective explorability. Remarkably, the gain that reflects the presence of a global random teleportation term remains consistently greater for simplicial complex structures across all analyzed values of $\delta$, see Fig. \ref{fig:gain_simplex_graph_summary} panel (c).
The gain is particularly pronounced in the region close to $p_1 = 1$, where the simplicial complex topology tends to form more elongated structures that benefit more of the random teleportation.

\section{Conclusion}
Summing up, we have here presented a natural formulation of a random walk process on simplicial complexes providing a simple yet conceptually fundamental model to describe cross-dimensional dynamics. This is achieved by introducing a novel operator that models hopping across multi-dimensional structures.  The ensuing stationary solution unveils the hidden ranking of  composing simplices of different order and kind. We further applied this framework to compute the simplicial complex explorability, restricted to  nodes, in presence of a long range stochastic relocation term. The effect of noise appears extremely beneficial when the walk is taken on higher order structures, as compared to what it happens on the corresponding binary graph. The codes employed are made publicly available in the repository \footnote{\url{https://github.com/diegofebbe/Random_walk_on_simplicial_complexes/tree/master}}.

\begin{acknowledgments}
\end{acknowledgments}

\bibliographystyle{apsrev4-2}
\bibliography{bibliography}

\appendix
\section{Long-range cross-dimensional random walk.}
\label{sec:appendix}
The process described by Eq.~\eqref{eq:pt} defines a random walk on a simplicial complex in which transitions are allowed only between incident simplices whose dimensions differ by one. Such a dynamics provides a simple, yet conceptually fundamental example of cross-dimensional diffusion over higher-order structures. Nevertheless, the same framework can be extended so as to include long-range jumps between simplices of non-adjacent dimensions, provided they are related by inclusion. More precisely, the walker can hop from simplex $i$ to simplex $j$, or vice versa, iff $\sigma_i^{(m)} \subset \sigma_j^{(n)}$ with $m<n.$\\
To encode such transitions we introduce the \emph{containment matrix}, whose entries are defined as
\begin{equation}
\label{eq:Cmn_def}
C_{m,n}(\sigma_i^{(m)},\sigma_j^{(n)})=
\begin{cases}
1, & \text{if } \sigma_i^{(m)}\subset \sigma_j^{(n)},\\
0, & \text{otherwise}.
\end{cases}
\end{equation}
that can be expressed in terms of the unsigned incidence matrices
\begin{equation}
\label{eq:Cmn_from_A}
C_{m,n}(i,j)=\mathbbm{1}\!\left({A}_{m+1}{A}_{m+2}\cdots {A}_n\right)_{ij},
\end{equation}
where $\mathbbm{1}(\cdot)$ denotes the entry-wise indicator function detecting the existence of at least one chain of nested simplices connecting $i$ to $j$, where to lighten the notations we have replaced $\sigma^{(m)}_i$ and $\sigma^{(n)}_j$ by $i$ and $j$.\\
By using the matrices $C_{m,n}$, we can define a global block matrix $\mathbf{C}\in\mathbb{R}^{N\times N}$ encapsulating all terms describing the transitions among simplices:
\begin{equation}
\label{eq:bigC}
\mathbf{C}=
\left(
\begin{matrix}
{O}_0 & {C}_{0,1} & {C}_{0,2} & \cdots & {C}_{0,D}\\
{C}_{0,1}^{\top} & {O}_1 & {C}_{1,2} & \cdots & {C}_{1,D}\\
{C}_{0,2}^{\top} & {C}_{1,2}^{\top} & {O}_2 & \cdots & {C}_{2,D}\\
\vdots & \vdots & \vdots & \ddots & \vdots\\
{C}_{0,D}^{\top} & {C}_{1,D}^{\top} & {C}_{2,D}^{\top} & \cdots & {O}_D
\end{matrix}
\right).
\end{equation}
To row-normalize such an operator we introduce the lower long-range generalized degree of each given simplex $\sigma_i^{(k)}$ as:
\begin{equation}
\label{eq:kappa_def}
\begin{aligned}
\mathfrak{d}_i^{(k)}=&
\sum_{m=0}^{k-1}\sum_{j=1}^{N_m} C_{m,k}(\sigma_j^{(m)},\sigma_i^{(k)})\\=&\sum_{m=0}^{k-1}\binom{k+1}{m+1}
=
2^{k+1}-2,
\end{aligned}
\end{equation}
and the upper long-range generalized degree
\begin{equation}
\label{eq:rho_def}
\mathsf{d}_i^{(k)}=
\sum_{n=k+1}^{D}\sum_{j=1}^{N_n} C_{k,n}(\sigma_i^{(k)},\sigma_j^{(n)}).
\end{equation}
Let now $\mathbb{D}\in\mathbb{R}^{N\times N}$ be the diagonal matrix whose diagonal entries are the generalized long-range degrees $\mathfrak{d}_i^{(k)} + \mathsf{d}_i^{(k)}$. Then, the transition matrix of the multi-dimensional random walk, generalizing the operator $\mathbf{M}$ defined in Eq.~\eqref{eq:MatrixM}, is defined by
\begin{equation}
\label{eq:bigM_long}
\mathbb{M}=\mathbb{D}^{-1}\mathbf{C},
\end{equation}
while the probability evolution reads
\begin{equation}
\label{eq:pt_long}
\vec{p}_{t+1}=\vec{p}_t\mathbb{M}.
\end{equation}

\end{document}